\documentclass[letter]{aa}

\usepackage[utf8]{inputenc}
\usepackage[varg]{txfonts}
\usepackage{textcomp}      

\usepackage{natbib}
\bibpunct{(}{)}{;}{a}{}{,} 
\usepackage{graphicx}   
\usepackage{epstopdf}
\usepackage{hyperref}   
\usepackage{longtable}
\usepackage{xcolor}
\usepackage{amsfonts,amssymb,latexsym}
\usepackage[normalem]{ulem}

\hypersetup{
    bookmarks=true,         
    colorlinks=true,       
    linkcolor=blue,          
    citecolor=blue,        
    filecolor=blue,      
   urlcolor=blue           
}

\def\ga{\,\,\raise0.14em\hbox{$>$}\kern-0.76em\lower0.28em\hbox{$\sim$}\,\,}
\def\la{\,\,\raise0.14em\hbox{$<$}\kern-0.76em\lower0.28em\hbox{$\sim$}\,\,}

\def\iso#1{$^{#1}$}
\def\Msun{$M_{\odot}$}

\def\cm3{cm$^{-3}$}

\def\chem#1#2{$\mathrm{^{#2}\kern-0.8pt#1}$}
\def\reac#1#2#3#4#5#6{$\mathrm{\, ^{#2}\kern-0.8pt{#1}\, ({#3}\, ,{#4})\, {}^{#6}\kern-0.8pt{#5}\, }$}

\def\be{\begin{equation}} 
\def\ee{\end{equation}}
\def\beqy{\begin{eqnarray}}
\def\eeqy{\end{eqnarray}}
\def\bmlet{\begin{mathletters}}
\def\emlet{\end{mathletters}}

\begin{document}

\title{Synthesis of thorium and uranium in asymptotic giant branch stars}

\author{A. Choplin   
\and S. Goriely
\and L. Siess}
\offprints{arthur.choplin@ulb.be}

\institute{
Institut d'Astronomie et d'Astrophysique, Universit\'e Libre de Bruxelles (ULB),  CP 226, B-1050 Brussels, Belgium
}

\date{Received --; accepted --}

\abstract
{
The intermediate neutron capture process (i-process) operates at neutron densities between those of the slow and rapid neutron-capture processes. It is believed to be triggered by the ingestion of protons in a convective helium-burning region. One possible astrophysical site is low-mass low-metallicity asymptotic giant branch (AGB) stars.
}
{
Although it has been widely believed that actinides, and most particularly Th and U, are exclusively produced by explosive r-process nucleosynthesis, we study here the possibility that actinides may also be significantly synthesized through i-process nucleosynthesis in AGB stars.
}
{
We computed a 1~\Msun\, model at [Fe/H]~$= -2.5$ with the stellar evolution code {\sf STAREVOL}. We used a nuclear network of 1160 species from H to Cf coupled to the transport processes. 
Models of various resolutions (temporal and spatial) that use different nuclear datasets are also considered for the analysis.
}
{
During the proton ingestion event, the neutron density in our AGB model goes up to $\sim 10^{15}$~cm$^{-3}$ and is shown to be high enough to give rise to the production of actinides. While most of the nuclear flow cycles in the neutron-rich Pb-Bi-Po region, a non-negligible fraction leaks towards heavier elements and eventually synthesizes actinides. 
The surface enrichment in Th and U is subject to nuclear and astrophysical model uncertainties that could be lowered in the future, in particular by a detailed analysis of the nuclear inputs that affect the neutron capture rates of neutron-rich isotopes between Pb and Pa, along the i-process path. 
One stellar candidate that may confirm the production of actinides by the i-process is the carbon-enhanced metal-poor (CEMP) r/s star \object{RAVE J094921.8-161722}, which shows Th lines in its spectrum. Its surface abundance is shown to be reasonably well reproduced by our AGB model, though abundances of light $N\simeq 50$ elements remain underestimated. Combined with cosmochronometry, this finding opens the way to dating the i-process event and thus obtaining a lower limit for the age of CEMP-r/s stars.
Such a dating is expected to be accurate only if surface abundances of Th and U can be extracted simultaneously.
}
{
We show that actinides can be synthesized in low-metallicity low-mass AGB stars through the i-process. This astrophysical site therefore potentially contributes to the Galactic enrichment of Th and U, which demonstrates that the r-process may not be the sole mechanism for the production of U and Th.
}


\titlerunning{Synthesis of Thorium and Uranium in AGB stars}

\authorrunning{A. Choplin et al. }

\maketitle


\section{Introduction}
\label{sect:intro}

The origin of elements heavier than iron is still debated \citep[e.g.][]{arnould20}. 
The slow (s-) and rapid (r-) neutron capture processes are thought to be responsible for the synthesis of most of these elements. It is classically accepted that the s-process ends in the Pb-Bi region with a series of fast $\alpha$ decays \citep{clayton67}, while the r-process can produce the heaviest nuclei, including the actinides. Up to now, the r-process alone has been held responsible for the nucleosynthesis of Th and U in the Universe.

Additionally, the intermediate neutron capture process, or i-process, with neutron densities between those of the the s- and r-processes ($N_n \simeq 10^{13} - 10^{15}$~cm$^{-3}$), was first proposed by Cowan \& Rose (1977) but has only recently been revived.
Different possible observational signatures of i-process nucleosynthesis were reported in carbon-enhanced metal-poor (CEMP) r/s stars \citep{jonsell06, lugaro12,dardelet14,roederer16,karinkuzhi21}, in barium anomalies in open clusters \citep{mishenina15}, and in the isotopic composition of pre-solar grains \citep{jadhav13,fujiya13,liu14}.
The i-process nucleosynthesis arises when protons are mixed into a convective helium-burning zone. 
The astrophysical site(s) hosting proton ingestion events (PIEs), and hence the i-process, is (are) still debated \citep[see e.g.][for a detailed list]{choplin21}.
One possibility is the early thermally pulsing (TP) phase of low-mass low-metallicity asymptotic giant branch (AGB) stars \citep[e.g.][]{iwamoto04,cristallo09b,stancliffe11,choplin21,goriely21}.

In this Letter we show that the i-process, expected to take place in low-metallicity AGB stars, can pass the Pb-Bi region and also synthesize actinides, including Th and U.
Section~\ref{sect:ing} presents the physical ingredients and models. 
Section~\ref{sect:nuc} focuses on the nucleosynthesis of actinides.
In Sect.~\ref{sect:comp} we compare our results with the abundances of \object{RAVE J094921.8-161722}, a CEMP r/s star that shows Th lines. Conclusions are given in Sect.~\ref{sect:concl}.

\begin{table*}[h!]
\scriptsize{
\caption{Main characteristics of the five models considered in this work. Given are the model label, nuclear dataset used (see text for details), spatial resolution parameter, $\epsilon_{\rm max}$ (see text for details), temporal resolution parameter, $\alpha$ (see text for details), maximum neutron density, surface $\log\epsilon$(Pb), $\log\epsilon$(Th), and $\log\epsilon$(U) 
after the PIE (all unstable actinides were decayed except $^{232}$Th, $^{235}$U, and $^{238}$U).
\label{table:1}
}
\begin{center}
\resizebox{17cm}{!} {
\begin{tabular}{lccccccc} 
\hline
Model &  Nuclear  &   $\epsilon_{\rm max}$    & $\alpha$ &   $\log(N_{\rm n,max}$)   & $\log\epsilon$(Pb) &  $\log\epsilon$(Th) & $\log\epsilon$(U)   \\
 label  &    dataset &     &      &       &                    \\
\hline
MA$\epsilon 8 \alpha 08$     &   A   &   0.08  & 0.008   & 15.33  &  3.78 & $0.74$ & $0.50$    \\
MA$\epsilon 4 \alpha 08$     &   A   &  0.04  & 0.008  & 15.31    & 3.88  & $0.90$ & $0.76$   \\
MA$\epsilon 4 \alpha 04$     &   A   &  0.04   & 0.004 & 15.34    &  3.83 & $0.78$ & $0.46$   \\
MB$\epsilon 4 \alpha 04$     &   B   & 0.04  & 0.004    & 15.34   &  3.84  & $0.16$ & $-0.26$  \\
MC$\epsilon 4 \alpha 04$     &   C   & 0.04  & 0.004    & 15.17   & 3.83   & $0.89$ & $0.82$  \\
\hline
\end{tabular}
}
\end{center}
}
\end{table*}


\section{The i-process model: Physical ingredients}
\label{sect:ing}

We considered the i-process taking place in a low-metallicity AGB star at the time of the proton ingestion, as described in detail in \cite{choplin21} and \cite{goriely21} and using the same physical ingredients. We briefly recall some important aspects.

The models are computed with the stellar evolution code  {\sf STAREVOL} \citep[][and references therein]{siess00, siess06, goriely18c}. 
We used the mass-loss rate from \cite{reimers75} from the main sequence up to the beginning of the AGB phase and then switched to the \cite{vassiliadis93} rate.
When the star becomes carbon rich, the opacity change due to the formation of molecules is included \citep{marigo02}. 
We used a mixing length parameter $\alpha = 1.75$, and no extra mixing (e.g. overshoot or thermohaline) was included. 
In our models, during a PIE, the transport and burning of chemicals are coupled. This means that the nucleosynthesis and transport equations are solved simultaneously once the structure has converged. 

For the relatively low neutron densities that characterize s-process nucleosynthesis (i.e. $N_n \la 10^{13}$~cm$^{-3}$), a network of 411 isotopes is traditionally used in {\sf STAREVOL}.
However, to reliably follow the i-process flow resulting from a PIE with $N_n > 10^{13}$~cm$^{-3}$, a larger network needs to be considered. We used a reaction network composed of 1160 nuclei from neutrons up to $^{253}$Cf and linked through 2123 nuclear reactions and decays. 
Nuclear reaction rates were taken from the Nuclear Astrophysics Library of the Universit\'e Libre de  Bruxelles\footnote{Available at http://www.astro.ulb.ac.be/bruslib} \citep{arnould06} and the updated experimental and theoretical rates from the NETGEN interface \citep{Xu13}.
The latest decay rates were taken from the recent NUBASE2020 release\footnote{https://www-nds.iaea.org/amdc/} \citep{kondev21}. 
It should be noted that fission processes are not included in the reaction network. 
Despite the fact that the i-process takes place on timescales of the order of days, the so-produced neutron-rich nuclei with $Z < 96$ have a significantly longer half-life against spontaneous fission, so spontaneous fission is not expected to affect the present results. 
For some isotopes with $Z \geq 94$, the timescale for neutron-induced fission can become comparable to that of the radiative neutron capture $(n,\gamma)$,  hindering the potential production of heavier actinides. However, on the basis of experimentally known cross-sections, along the i-process path, only $^{241}$Pu appears to have  an ($n,f$) cross-section larger than its $(n,\gamma)$ cross-section (by a factor of about 4 at the thermal neutron energy of about 20~keV). Taking into account neutron-induced fission may affect the quantitative abundance estimates, but it will not change the fact that actinides are produced. For this reason, future simulations may need to include neutron-induced fission processes in the network.

 \begin{figure*}[t]
\includegraphics[scale=0.7, trim = 0cm 0cm 0cm 0.cm]{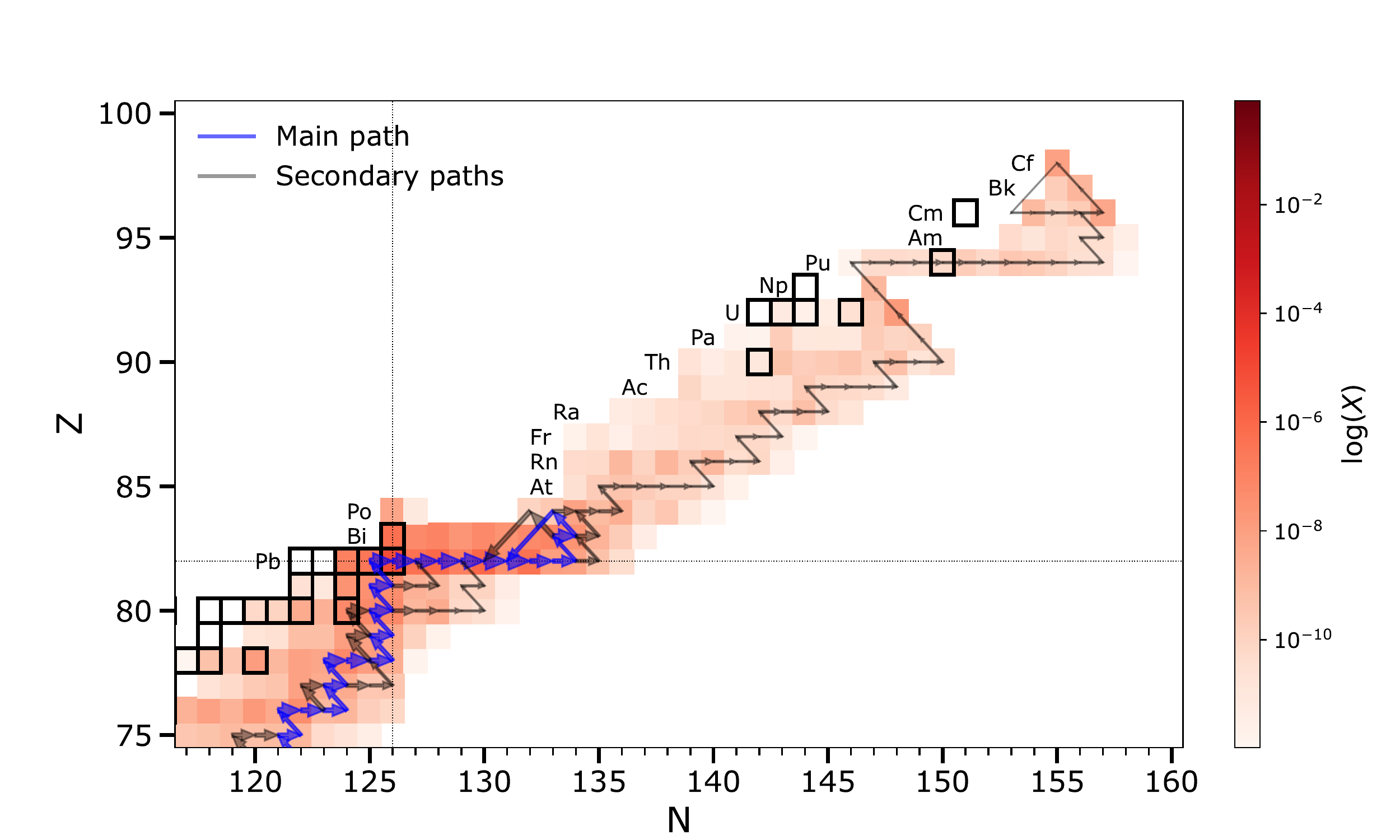}
\caption{
Main (blue) and secondary (black) i-process paths (starting from $^{56}$Fe) in the MA$\epsilon 4 \alpha 04$ model, at the bottom of the convective thermal pulse, at the time of maximum neutron density ($N_n = 2.17 \times 10^{15}$~cm$^{-3}$).
A secondary path is considered as such if at least 30~\% of the total flux goes through it. 
The size of the arrows scales with the flux. 
The black squares highlight the stable and long-lived isotopes \iso{232}Th, \iso{234}U, \iso{235}U, \iso{236}U, \iso{238}U, \iso{237}Np, \iso{244}Pu, and \iso{247}Cm. 
The colour of the different nuclei corresponds to their mass fraction at that time.
}
\label{fig:path}
\end{figure*}

In this Letter we focus on an AGB model of 1~\Msun\ with [Fe/H]~$=-2.5$. As extensively discussed in \cite{choplin21}, this model experiences a PIE during the early TP-AGB phase. 
During the PIE, five different models were computed with different spatial discretization, temporal resolution, and nuclear datasets. The spatial and temporal resolutions are controlled by the $\epsilon_{\rm max}$ and $\alpha$ parameters, respectively. 
More specifically, a new mesh point was added if the relative variation in a structural variable exceeds the threshold value, $\epsilon_\mathrm{max}$, between two adjacent shells.
The parameter $\alpha$ controls by how much the timestep is allowed to change based on the relative variations in the structure variables. More details are given in Sects.~4.1 and 4.2 of \cite{choplin21}. 
The lower these parameters, the higher the resolution. In addition to such astrophysical uncertainties, the abundance calculation is also affected by nuclear uncertainties, in particular due to the large amount (about 70\%) of unmeasured neutron capture rates involved during i-process nucleosynthesis. To study nuclear uncertainties, as in \citet{goriely21}, different sets of theoretical $(n,\gamma)$ rates were considered, and their impact on the abundances was obtained by consistently re-calculating the full PIE event with {\sf STAREVOL} and the updated nuclear network. More specifically, we considered three sets of neutron capture rates. In addition to the fiducial rates (model A), we also performed the calculations with model B -- which uses the photon strength functions from \citet{Goriely04} instead of \citet{Goriely19}, as included in model A -- and model C, which adds to model A the contribution from the direct capture component \citep{Sieja21}. Models B and C were chosen out of the various sets included in \citet{goriely21} because they give rise to lower and upper limits to actinide production, as discussed below.   
Table~\ref{table:1} reports the characteristics of these five models.

 \begin{figure*}[t]
\includegraphics[scale=0.55, trim = 2cm 0cm 3cm 0cm]{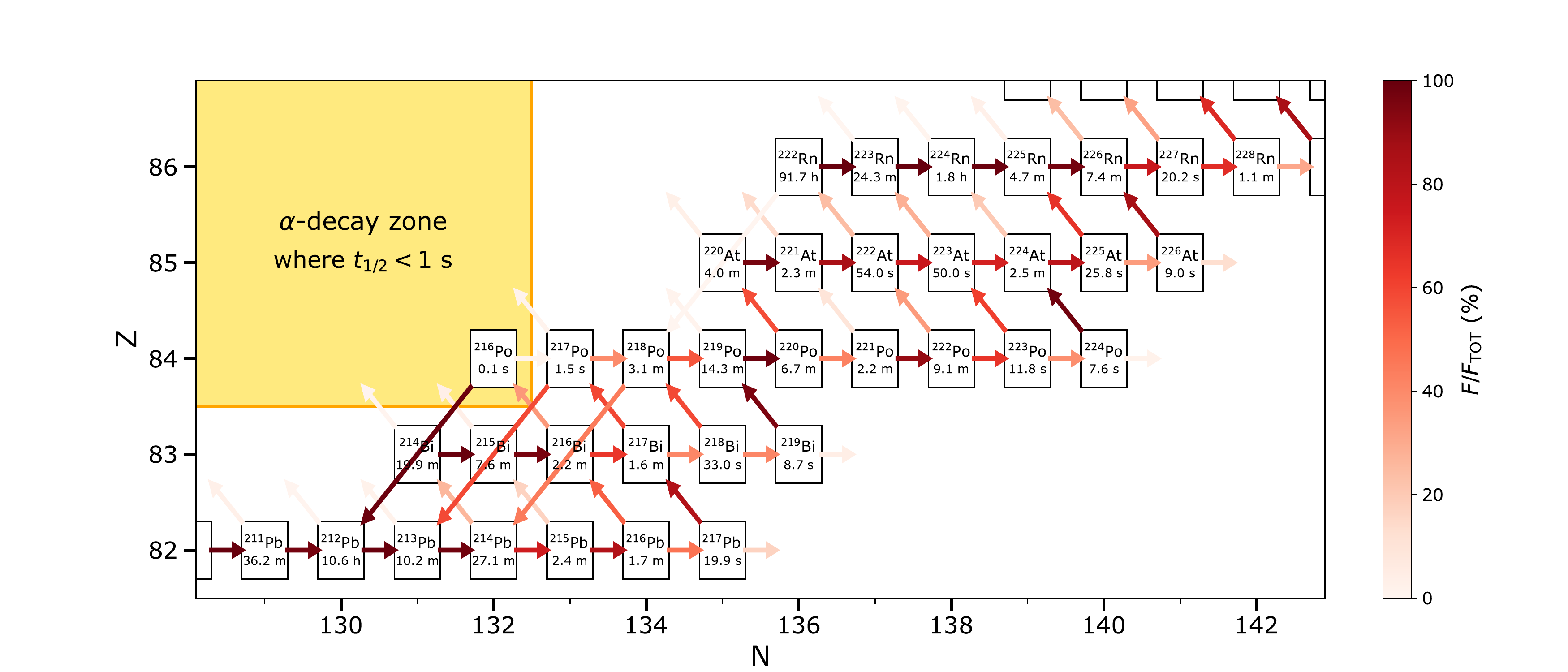}
\caption{Flow chart in the Pb-Rn region in the MA$\epsilon 4 \alpha 04$ model, at maximum neutron density ($N_n = 2.17 \times 10^{15}$~cm$^{-3}$). 
The coloured arrows from a given isotope, $i$, represent the flux ratio $F/F_{\rm tot}$, where $F$ corresponds to the flux of the $(n,\gamma)$ reaction $F_{\rm (n,\gamma)} = N_{\rm av}  \rho Y_n Y_i \langle \sigma v \rangle$, $\beta$ decay $F_{\rm \beta} = \lambda_\beta Y_i$, or  $\alpha$ decay $F_{\rm \alpha} = \lambda_\alpha Y_i$, and $F_{\rm tot} = F_{\rm (n,\gamma)} + F_{\rm \beta} + F_{\rm \alpha}$ ($Y_n$ and $Y_i$ being the molar mass fraction of the neutrons and target, respectively, $\langle \sigma v \rangle$ the nuclear reaction rate, and $\lambda_\beta$ and $\lambda_\alpha$ the decay rates).
The yellow area shows the nuclei experiencing $\alpha$ decays with $t_{1/2} < 1$~s.
}
\label{fig:flux}
\end{figure*}

\section{Synthesis of actinides}
\label{sect:nuc}

Proton ingestion events that lead to i-process nucleosynthesis can have a profound impact on AGB structure and evolution. 
This was extensively discussed in \cite{choplin21,choplin22}. 
In this Letter we focus on the possible synthesis of the heaviest elements beyond Pb through i-process nucleosynthesis. 

In the fiducial MA$\epsilon 4 \alpha 04$ model, at the neutron density peak, the i-process mainly follows the blue path shown in Fig~\ref{fig:path}. From $^{209}$Pb to $^{215}$Pb, the timescale against neutron capture, $\tau_n$, is significantly smaller than the $\beta$-decay timescale, $\tau_{\beta}$, such that the ($n,\gamma$) channel dominates. 
At $^{216}$Pb, $\beta^-$ somewhat dominates and the main path follows the chain $^{216}$Pb($\beta^-$)$^{216}$Bi($n,\gamma$)$^{217}$Bi($\beta^-$)$^{217}$Po($\gamma,\alpha$)$^{213}$Pb, which eventually forms a loop.

However, at  $^{216}$Pb, $^{217}$Bi, and $^{217}$Po, the ($n,\gamma$) reactions compete with the $\beta$ decays, and  a significant fraction of the flux (at least 30~\%) goes into a secondary path (black paths in Fig~\ref{fig:path}).
These branching points are important since they determine if the flux cycles in the Pb-Bi-Po region or continues up to heavier elements. 
At $^{216}$Pb, the timescales against $\beta^-$ decay and neutron capture (at $T = 250$~MK and maximum $N_n$) are $\tau_{\beta} = 2.5$~min and $\tau_n = 2.7$~min, respectively. Although the $\beta^{-}$ channel dominates, more than 30~\% of the flux continues to $^{217}$Pb.
This is similar for $^{217}$Bi ($\tau_{\beta} = 2.3$~min and $\tau_n  = 3.4$~min) and $^{217}$Po ($\tau_{\beta} = 2.2$~s and $\tau_n  = 3.4$~s).

A crucial condition for synthesizing actinides is to have a neutron density high enough to pass the extremely fast $\alpha$-decay region at $Z\ge 84$ and $126 \le N \le 132$, which inevitably brings the nuclear flow back to the Pb region (yellow area in Fig.~\ref{fig:flux}). 
If the neutron flux is too low, isotopes with $N > 134$ will not form and the nucleosynthesis flow will cycle in the Pb-Bi-Po region with $N<134$.
The minimum neutron density needed to overtake this fast $\alpha$-decay zone and build up actinides is about $10^{15}$~cm$^{-3}$, though the determination of its exact value is still hindered by the unknown reaction rates of the neutron-rich Pb-Po isotopes involved. As reported in Table~\ref{table:1}, all our i-process simulations in the 1~\Msun\ [Fe/H]~$=-2.5$ model star lead to a significant enrichment of the stellar surface in Th and U. Our calculations also indicate a strong correlation between Pb and actinide abundances. We thus expect a CEMP r/s star enriched in Th and U to also have a high Pb abundance.

When inspecting the results for the different nuclear datasets, a few key reactions show important differences between the nuclear models. 
At $^{217}$Po, for instance (which is on the main path in Fig.~\ref{fig:path}), 29, 3, and 26~\% of the flux follows the $(n,\gamma)$ channel for models A, B, and C, respectively. 
Almost all the remaining flux follows the $\alpha$-decay channel back towards $^{213}$Pb.
This contributes to lowering the production of actinides in model B and reduces the Th and U surface abundances after the PIE (Table~\ref{table:1}).
We report in Table~\ref{table:2} the most uncertain, and hence critical, $(n,\gamma)$ reactions of relevance according to nuclear models A, B, and C.
Only the relevant reactions for i-process nucleosynthesis starting from Pb are mentioned.
An accurate determination of these reaction rates is needed to reduce the uncertainties regarding the production of actinides in low-metallicity AGB stars.

The production of actinides is mostly unchanged under different spatial and temporal discretizations during the PIE. After the PIE, the surface\footnote{$\log\epsilon(X) = \log_{10} (N_{\rm X} / N_{\rm H}) + 12$, where $N_{\rm X}$ and $N_{\rm H}$ refer
to the numbers of atoms of elements X and hydrogen, respectively.} $\log\epsilon$(Th) shows a scatter of 0.16 dex and the surface $\log\epsilon$(U) a scatter of 0.3 dex (cf. the first three models in Table~\ref{table:1}). 
We notice that a higher spatial resolution slightly increases the production of Th and U, while a better temporal resolution reduces the production.

\section{Comparison to the CEMP r/s star RAVE J094921.8-161722}
\label{sect:comp}

\cite{gull18} reported the discovery of \object{RAVE J094921.8-161722} (hereafter J0949-1617), a red giant, CEMP star with ${\rm [Fe/H]}=-2.2$. 
This star has a chemical composition between that of the s- and r-processes and was interpreted as having been polluted by both processes. 
The only argument for justifying pollution by the r-process is based on the presence of Th in the spectrum, with an abundance estimated to $\log \epsilon (\mathrm{Th}) = -1.70 \pm 0.20$. 
The scenario proposed by \cite{gull18} corresponds to a star formed in a zone enriched by a prior r-process event that later accreted some s-rich material from an AGB companion. Its chemical abundances would then reflect a combination of s- and r-processes. 
In this section we discuss the alternative possibility of pollution by a low-metallicity AGB companion that experienced i-process nucleosynthesis.

To explain the present composition of J0949-1617, we freely varied the dilution factor, $f$, for each of the five models so as to minimize the $\chi_{\nu}^2$ between theory and observation 
\citep[cf.][for details]{choplin21,choplin21cor}.
The $\chi_{\nu}^2$ was estimated based on the $\log \epsilon$ abundance of all available elements with $Z>30,$ except Th because its abundance decreased between the time of pollution and the observation. 

\begin{figure*}[t]
\includegraphics[scale=0.7, trim = 2cm 0cm 0cm 0.cm]{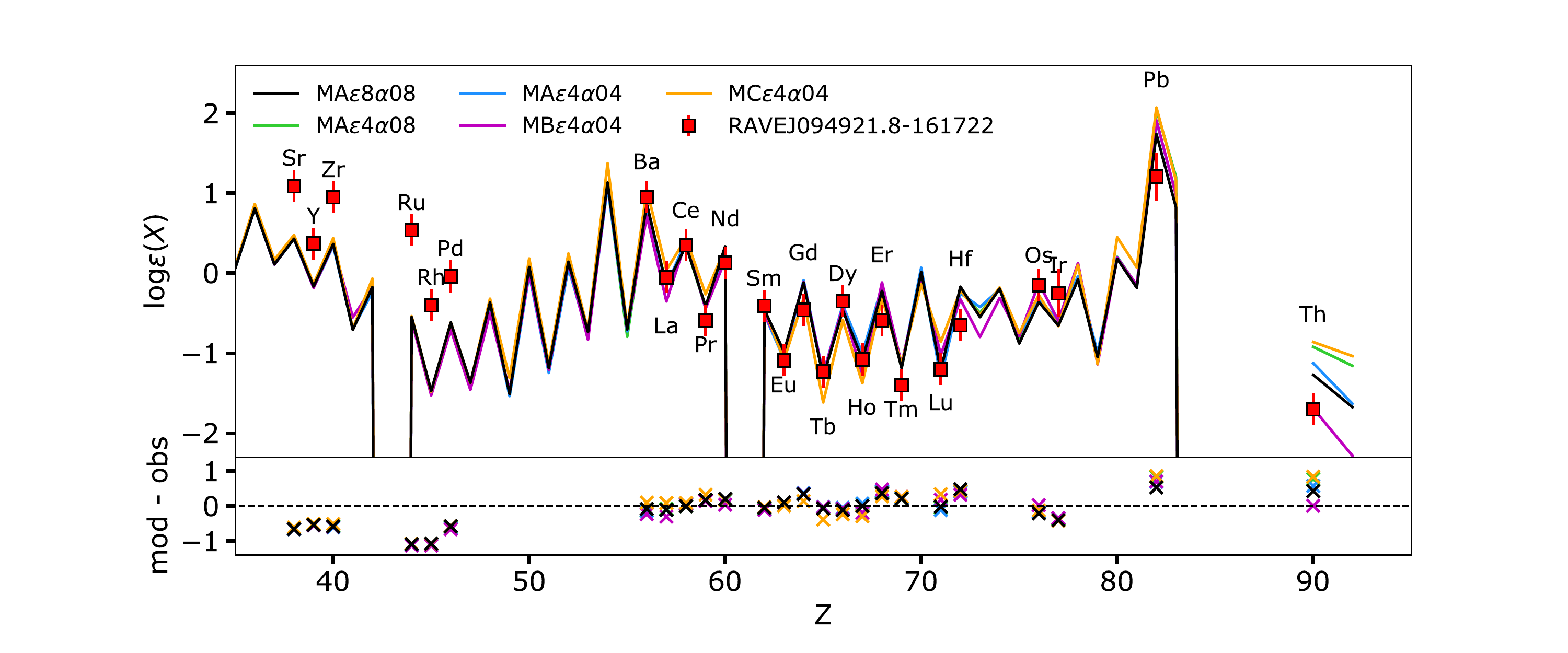}
\caption{ Best fits to the abundances of the CEMP r/s star \object{RAVE J094921.8-161722} \citep{gull18} using the five AGB models listed in Table~\ref{table:1}. 
The abundance of uranium is based on the \iso{238}U isotope only; the \iso{235}U isotope, with a half-life of $t_{1/2}=0.7$~Gyr, was assumed to have fully decayed. }
\label{fig:res}
\end{figure*}

The five best fits and residuals are shown in Fig.~\ref{fig:res}. 
The dilution factors are $0.98<f<0.99$ for all models, meaning that $1-2$~\% of AGB material is mixed with $98-99$~\% of interstellar medium material. 
The agreement between J0949-1617 and the models is reasonably good for elements with $55<Z<80$ (residuals are less than $0.5$~dex), as well as for Sr, Y, Zr, and Pd (although these elements are underproduced by 0.5 dex). 
The elements Ru and Rh are systematically underproduced by 1~dex, and Pb is overproduced by $0.5-1$~dex. 
But more importantly, the predictions for the s elements (e.g. La and Ba) and r elements (e.g. Eu) in the $N\simeq 82$ region are in excellent agreement with observations.

In principle, the time since the i-process event can be estimated from the Th/Eu cosmo\-chronometry through the expression 
$\Delta t = 46.67  \, [ \,  \log (\mathrm{Th/Eu})_{\rm AGB} - \log (\mathrm{Th/Eu})_{\rm now} \, ],$ where\footnote{We note that $\log (\mathrm{X/Y}) = \log \epsilon (\mathrm{X}) - \log \epsilon (\mathrm{Y})$.} $\log (\mathrm{Th/Eu})_{\rm now} = -0.61$ \citep{Fowler60,cayrel01}.
This gives a lower limit on the age of J0949-1617. 
The age estimate for each of our five models is reported in Table~\ref{table:3}. 
The scatter of 0.92~dex in $\log (\mathrm{Th/Eu})_{\rm AGB}$ leads to an uncertainty of 43~Gyr on the `age' of J0949-1617. This confirms that an age estimate based on the Th/Eu ratio is not realistic. Even with an extreme precision of 0.1~dex on $\log (\mathrm{Th/Eu})_{\rm AGB}$, the age uncertainty would be about 5~Gyr.  
A better age estimate can be obtained using the Th/U ratio instead. 
In this case, the age is derived through the expression $\Delta t' = 21.76  \, [ \,  \log (\mathrm{Th/U})_{\rm AGB} - \log (\mathrm{Th/U})_{\rm now} \, ]$. 
Since U has not been determined at the surface of J0949-1617, $\log (\mathrm{Th/U})_{\rm now}$ is not available. However, the age uncertainty based on the Th/U ratio can be estimated by comparing the values of $\Delta t'$ between two models. 
The age difference, $\delta(\Delta t')$, between the considered model and the reference model, MA$\epsilon 4 \alpha 04$, is reported in the last column of Table~\ref{table:3}. It is given by 
$\delta(\Delta t') = 21.76  \, [ \,  \log (\mathrm{Th/U})_{\rm AGB} - \log (\mathrm{Th/U})_{\rm AGB\_REF} \, ]$. 

Different spatial and temporal discretizations lead to an age uncertainty, $\delta(\Delta t')$, of $\sim 6$~Gyr and different nuclear physics to an uncertainty of $\sim 9$~Gyr.
Although such an uncertainty remains large, an improved determination of the theoretical reaction rates above Pb along the i-process path (especially for the reactions reported in Table~\ref{table:2})  would lower the impact of the nuclear uncertainty on the predicted Th and U abundances.

\begin{table}[t]
\scriptsize{
\caption{ 
Most uncertain $(n,\gamma)$ reactions (above Pb) relevant to i-process nucleosynthesis, according to the three different nuclear models, A, B, and C. 
The numbers give the ratios of the Maxwellian-averaged cross-section at $T=250$~MK, which corresponds to the temperature at the bottom of the thermal pulse during the PIE.
This table includes all $(n,\gamma)$ reactions that differ by at least of factor of 10 between two different nuclear models.
\label{table:2}
}
\begin{center}
\resizebox{8cm}{!} {
\begin{tabular}{lcc} 
\hline
  &   $\langle \sigma \rangle_{\rm B} / \langle \sigma \rangle_{\rm A}$    & $\langle \sigma \rangle_{\rm C} / \langle \sigma \rangle_{\rm A}$        \\
\hline
$^{216}$Pb$(n,\gamma)$     &   1.57e-2  & 9.71e-1  \\
$^{217}$Po$(n,\gamma)$     &   8.72e-2  & 8.45e-2  \\
$^{229}$Rn$(n,\gamma)$     &   9.85e+0  & 1.21e+1  \\
$^{224}$Fr$(n,\gamma)$     &   9.63e+0  & 1.23e+1  \\
$^{230}$Fr$(n,\gamma)$     &   5.68e+0  & 3.79e+1  \\
$^{228}$Ra$(n,\gamma)$     &   8.93e-1  & 8.37e-2  \\
$^{232}$Ra$(n,\gamma)$     &   1.52e+1  & 7.64e+0  \\
$^{233}$Ra$(n,\gamma)$     &   1.29e+1  & 6.80e+0  \\
$^{228}$Ac$(n,\gamma)$     &   5.69e-1  & 8.96e+1  \\
$^{236}$Ac$(n,\gamma)$     &   3.39e+0  & 1.38e+1  \\
$^{232}$Pa$(n,\gamma)$     &   3.61e+0  & 1.13e+1  \\
$^{234}$Pa$(n,\gamma)$     &   3.62e+1  & 2.56e+1  \\
$^{238}$Pa$(n,\gamma)$     &   1.33e+1  & 3.08e+1  \\
\hline
\end{tabular}
}
\end{center}
}
\end{table}


\begin{table*}[t]
\caption{Abundances of the best-fit models to J0949-1617. We also give the `age' estimate ($\Delta t$, i.e. the time since the i-process took place, a lower limit to the stellar age; see text for details) of J0949-1617 based on the Th/Eu abundance ratio and the difference for the age estimate, $\delta (\Delta t')$, with respect to the MA$\epsilon 4 \alpha 04$ model, if relying on the Th/U abundance ratio (see text for details). Note that the age estimate for the MB$\epsilon 4 \alpha 04$ model is negative because the Th/Eu ratio in this model is lower than the presently observed ratio. The abundance of U is based solely on the \iso{238}U isotope since \iso{235}U, with a half-life of $t_{1/2}=0.7$~Gyr, was assumed to have fully decayed. 
}
\label{table:3}
\begin{center}
\resizebox{17.5cm}{!} {
\begin{tabular}{lcccccccccccc} 
\hline
Model  & $\log\epsilon$(Eu) & $\log\epsilon$(Th)  &   $\log\epsilon$(U) & $\log(\mathrm{Th/Eu})_{\rm AGB}$  & $\log(\mathrm{Th/U})_{\rm AGB}$ & $\Delta t$ & $\delta (\Delta t')$\\
 label &           &          &               &         & & [Gyr] & [Gyr] \\
\hline
MA$\epsilon 8 \alpha 08$     &   $-0.98$   &   $-1.27$  & $-1.68$   & $-0.29$ & $0.40$ & $14.95$  & $-2.25$ \\
MA$\epsilon 4 \alpha 08$     &   $-0.99$   &   $-0.92$  & $-1.16$   & $0.07$ & $0.24$ & $31.80$   & $-5.89$ \\
MA$\epsilon 4 \alpha 04$     &   $-1.08$   &   $-1.12$  & $-1.63$   & $-0.05$ & $0.51$ & $26.16$   & $0$ \\
MB$\epsilon 4 \alpha 04$     &   $-1.00$   &   $-1.69$  & $-2.28$   & $-0.69$ & $0.59$ & $-3.89$   & $+1.81$ \\
MC$\epsilon 4 \alpha 04$     &   $-1.09$   &   $-0.86$  & $-1.04$   & $0.23$ & $0.18$ & $39.01$   & $-7.21$ \\
\hline
\end{tabular}
}
\end{center}
\end{table*}


\section{Conclusions}
\label{sect:concl}

In this Letter we have shown that a low-metallicity ([Fe/H]~$=-2.5$) low-mass (1~\Msun) AGB star experiencing a successful i-process (through PIE) can synthesize actinides, including a significant amount of Th and U. This production of actinides is also expected to lead to a significant over-abundance of Pb.

While the main i-process path cycles in the neutron-rich Pb-Bi-Po region, a potentially non-negligible fraction of the flux can leak towards actinides.
A neutron density of about $10^{15}$~cm$^{-3}$ is necessary to bypass the fast $\alpha$-decay region and build up actinides.
The isotopes $^{216}$Pb, $^{217}$Bi, and $^{217}$Po are shown to be important branching points where neutron capture competes with beta decay.

Varying the spatial and temporal resolution during the PIE leads to an uncertainty of 0.16~dex on Th and 0.30~dex on U. 
The nuclear uncertainties have a larger impact, with about 0.7~dex for Th and 1.1~dex for U. 
The different nuclear models allow us to highlight the most uncertain rates above Pb that affect the actinide nucleosynthesis (Table~\ref{table:2}).

We compared our models to the abundances of the CEMP r/s star \object{RAVE J094921.8-161722}, in which Th was detected. We find a rather good agreement for $55<Z<80$ elements, showing that the surface composition is compatible with an i-process origin. Most importantly, the Th enrichment can now be explained by the i-process and does not require additional pollution by the r-process. 
Such a finding also opens the way to a possible estimate of the time since the i-process event, through actinide-based cosmochronometry, which would provide a lower limit on the age of the CEMP r/s star. 
An estimation via the Th/Eu ratio suffers very large uncertainties, and a more accurate (although still uncertain) age indicator could be obtained if the U abundance can be estimated in CEMP r/s stars.
An accurate spectroscopic determination of Th (and hopefully U) is needed, especially in view of the difficulties in estimating the surface Th (and U) abundances in carbon-rich stars blended by CH molecules.
However, for a reliable age estimate, astrophysical and nuclear uncertainties need to be reduced first.

Despite the remaining uncertainties affecting the models, it seems clear that the high neutron densities encountered in low-metallicity low-mass AGB stars through the i-process can be a source of actinides, including Th and U. We consequently now know that Th and U are not  exclusively produced by r-process nucleosynthesis.
The AGB mass and metallicity ranges within which actinides can be formed remain to be explored. 
Another open question is whether or not other astrophysical sites that host the i-process can synthesize actinides.

\section*{Acknowledgments}
This work was supported by the Fonds de la Recherche Scientifique-FNRS under Grant No IISN 4.4502.19. 
L.S. and S.G. are senior FRS-F.N.R.S. research associates.

\bibliographystyle{aa}
\bibliography{astro.bib}

\begin{thebibliography}{33}
\expandafter\ifx\csname natexlab\endcsname\relax\def\natexlab#1{#1}\fi

\bibitem[{Arnould \& Goriely(2006)}]{arnould06}
Arnould, M. \& Goriely, S. 2006, Nucl. Phys. A, 777, 157

\bibitem[{{Arnould} \& {Goriely}(2020)}]{arnould20}
{Arnould}, M. \& {Goriely}, S. 2020, Progress in Particle and Nuclear Physics,
  112, 103766

\bibitem[{{Cayrel} {et~al.}(2001){Cayrel}, {Hill}, {Beers}, {Barbuy}, {Spite},
  {Spite}, {Plez}, {Andersen}, {Bonifacio}, {Fran{\c{c}}ois}, {Molaro},
  {Nordstr{\"o}m}, \& {Primas}}]{cayrel01}
{Cayrel}, R., {Hill}, V., {Beers}, T.~C., {et~al.} 2001, \nat, 409, 691

\bibitem[{{Choplin} {et~al.}(2021){Choplin}, {Siess}, \& {Goriely}}]{choplin21}
{Choplin}, A., {Siess}, L., \& {Goriely}, S. 2021, \aap, 648, A119

\bibitem[{{Choplin} {et~al.}(2022{\natexlab{a}}){Choplin}, {Siess}, \&
  {Goriely}}]{choplin21cor}
{Choplin}, A., {Siess}, L., \& {Goriely}, S. 2022{\natexlab{a}}, \aap, 662, C3

\bibitem[{{Choplin} {et~al.}(2022{\natexlab{b}}){Choplin}, {Siess}, \&
  {Goriely}}]{choplin22}
{Choplin}, A., {Siess}, L., \& {Goriely}, S. 2022{\natexlab{b}}, arXiv
  e-prints, arXiv:2209.10303

\bibitem[{{Clayton} \& {Rassbach}(1967)}]{clayton67}
{Clayton}, D.~D. \& {Rassbach}, M.~E. 1967, \apj, 148, 69

\bibitem[{{Cristallo} {et~al.}(2009){Cristallo}, {Piersanti}, {Straniero},
  {Gallino}, {Dom{\'\i}nguez}, \& {K{\"a}ppeler}}]{cristallo09b}
{Cristallo}, S., {Piersanti}, L., {Straniero}, O., {et~al.} 2009, \pasa, 26,
  139

\bibitem[{{Dardelet} {et~al.}(2014){Dardelet}, {Ritter}, {Prado}, {Heringer},
  {Higgs}, {Sandalski}, {Jones}, {Denisenkov}, {Venn}, {Bertolli}, {Pignatari},
  {Woodward}, \& {Herwig}}]{dardelet14}
{Dardelet}, L., {Ritter}, C., {Prado}, P., {et~al.} 2014, in XIII Nuclei in the
  Cosmos (NIC XIII), 145

\bibitem[{Fowler \& Hoyle(1960)}]{Fowler60}
Fowler, W. \& Hoyle, F. 1960, Ann. Phys., 10, 280

\bibitem[{{Fujiya} {et~al.}(2013){Fujiya}, {Hoppe}, {Zinner}, {Pignatari}, \&
  {Herwig}}]{fujiya13}
{Fujiya}, W., {Hoppe}, P., {Zinner}, E., {Pignatari}, M., \& {Herwig}, F. 2013,
  \apjl, 776, L29

\bibitem[{Goriely {et~al.}(2019)Goriely, Dimitriou, Wiedeking, Belgya,
  Firestone, Kopecky, Krticka, Plujko, Schwengner, Siem, Utsunomiya, Hilaire,
  P{\'e}ru, Cho, Filipescu, Iwamoto, Kawano, Varlamov, \& Xu}]{Goriely19}
Goriely, S., Dimitriou, P., Wiedeking, M., {et~al.} 2019, Eur. Phys. J. A, 55,
  172

\bibitem[{Goriely {et~al.}(2004)Goriely, Khan, \& Samyn}]{Goriely04}
Goriely, S., Khan, E., \& Samyn, M. 2004, Nucl. Phys. A, 739, 331

\bibitem[{Goriely \& Siess(2018)}]{goriely18c}
Goriely, S. \& Siess, L. 2018, A\&A, 609, A29

\bibitem[{{Goriely} {et~al.}(2021){Goriely}, {Siess}, \& {Choplin}}]{goriely21}
{Goriely}, S., {Siess}, L., \& {Choplin}, A. 2021, \aap, 654, A129

\bibitem[{Gull {et~al.}(2018)Gull, Frebel, Cain, Placco, Ji, Abate, Ezzeddine,
  Karakas, Hansen, Sakari, Holmbeck, Santucci, Casey, \& Beers}]{gull18}
Gull, M., Frebel, A., Cain, M.~G., {et~al.} 2018, The Astrophysical Journal,
  862, 174

\bibitem[{{Iwamoto} {et~al.}(2004){Iwamoto}, {Kajino}, {Mathews}, {Fujimoto},
  \& {Aoki}}]{iwamoto04}
{Iwamoto}, N., {Kajino}, T., {Mathews}, G.~J., {Fujimoto}, M.~Y., \& {Aoki}, W.
  2004, \apj, 602, 377

\bibitem[{{Jadhav} {et~al.}(2013){Jadhav}, {Pignatari}, {Herwig}, {Zinner},
  {Gallino}, \& {Huss}}]{jadhav13}
{Jadhav}, M., {Pignatari}, M., {Herwig}, F., {et~al.} 2013, \apjl, 777, L27

\bibitem[{{Jonsell} {et~al.}(2006){Jonsell}, {Barklem}, {Gustafsson},
  {Christlieb}, {Hill}, {Beers}, \& {Holmberg}}]{jonsell06}
{Jonsell}, K., {Barklem}, P.~S., {Gustafsson}, B., {et~al.} 2006, \aap, 451,
  651

\bibitem[{{Karinkuzhi} {et~al.}(2021){Karinkuzhi}, {Van Eck}, {Goriely},
  {Siess}, {Jorissen}, {Merle}, {Escorza}, \& {Masseron}}]{karinkuzhi21}
{Karinkuzhi}, D., {Van Eck}, S., {Goriely}, S., {et~al.} 2021, \aap, 645, A61

\bibitem[{{Kondev} {et~al.}(2021){Kondev}, {Wang}, {Huang}, {Naimi}, \&
  {Audi}}]{kondev21}
{Kondev}, F.~G., {Wang}, M., {Huang}, W.~J., {Naimi}, S., \& {Audi}, G. 2021,
  Chinese Physics C, 45, 030001

\bibitem[{{Liu} {et~al.}(2014){Liu}, {Savina}, {Davis}, {Gallino}, {Straniero},
  {Gyngard}, {Pellin}, {Willingham}, {Dauphas}, {Pignatari}, {Bisterzo},
  {Cristallo}, \& {Herwig}}]{liu14}
{Liu}, N., {Savina}, M.~R., {Davis}, A.~M., {et~al.} 2014, \apj, 786, 66

\bibitem[{{Lugaro} {et~al.}(2012){Lugaro}, {Karakas}, {Stancliffe}, \&
  {Rijs}}]{lugaro12}
{Lugaro}, M., {Karakas}, A.~I., {Stancliffe}, R.~J., \& {Rijs}, C. 2012, \apj,
  747, 2

\bibitem[{{Marigo}(2002)}]{marigo02}
{Marigo}, P. 2002, \aap, 387, 507

\bibitem[{{Mishenina} {et~al.}(2015){Mishenina}, {Pignatari}, {Carraro},
  {Kovtyukh}, {Monaco}, {Korotin}, {Shereta}, {Yegorova}, \&
  {Herwig}}]{mishenina15}
{Mishenina}, T., {Pignatari}, M., {Carraro}, G., {et~al.} 2015, \mnras, 446,
  3651

\bibitem[{{Reimers}(1975)}]{reimers75}
{Reimers}, D. 1975, Memoires of the Societe Royale des Sciences de Liege, 8,
  369

\bibitem[{{Roederer} {et~al.}(2016){Roederer}, {Karakas}, {Pignatari}, \&
  {Herwig}}]{roederer16}
{Roederer}, I.~U., {Karakas}, A.~I., {Pignatari}, M., \& {Herwig}, F. 2016,
  \apj, 821, 37

\bibitem[{Sieja \& Goriely(2021)}]{Sieja21}
Sieja, K. \& Goriely, S. 2021, Eur. Phys. J. A, 57, 110

\bibitem[{{Siess}(2006)}]{siess06}
{Siess}, L. 2006, \aap, 448, 717

\bibitem[{Siess {et~al.}(2000)Siess, Dufour, \& Forestini}]{siess00}
Siess, L., Dufour, E., \& Forestini, M. 2000, A\&A, 358, 593

\bibitem[{{Stancliffe} {et~al.}(2011){Stancliffe}, {Dearborn}, {Lattanzio},
  {Heap}, \& {Campbell}}]{stancliffe11}
{Stancliffe}, R.~J., {Dearborn}, D. S.~P., {Lattanzio}, J.~C., {Heap}, S.~A.,
  \& {Campbell}, S.~W. 2011, \apj, 742, 121

\bibitem[{{Vassiliadis} \& {Wood}(1993)}]{vassiliadis93}
{Vassiliadis}, E. \& {Wood}, P.~R. 1993, \apj, 413, 641

\bibitem[{Xu {et~al.}(2013)Xu, Goriely, Jorissen, Chen, \& Arnould}]{Xu13}
Xu, Y., Goriely, S., Jorissen, A., Chen, G., \& Arnould, M. 2013, A\&A, 549, 10

\end{thebibliography}

\end{document}